\documentclass[american,english,format=acmsmall, review=false, screen=true,nonacm=true]{acmart}
\usepackage{graphicx} 

\usepackage{hyperref}
\hypersetup{
    colorlinks=true,
    linkcolor=blue,
    filecolor=magenta,      
    urlcolor=cyan,
    pdfpagemode=FullScreen,
    }

\usepackage{tikz}
\usetikzlibrary{decorations.pathmorphing,trees,positioning} 

\theoremstyle{definition}
\newtheorem{defn}{Definition}[section]
\theoremstyle{remark}
\newtheorem*{rem}{Remark}

\usepackage{tcolorbox}

\usepackage{varwidth}
\makeatletter
\newenvironment{cellvarwidth}[1][t]
    {\begin{varwidth}[#1]{\linewidth}}
    {\@finalstrut\@arstrutbox\end{varwidth}}
\makeatother

\usepackage{alltt}
\usepackage{url}

\title{Mathematical Modeling of a Cognitive Continuum Digital Shadow for Large-Scale, Cross-Facility Workflows}
\author{Mark Asch}
\orcid{0000-0002-1014-4097}
\affiliation{\institution{LAMFA, Université de Picardie Jules Verne}\streetaddress{33 rue Saint Leu}\city{Amiens}\postcode{80000}\country{France}}
\author{Marius Garénaux Gruau}
\affiliation{\institution{IRISA}\city	{Rennes}\postcode{35000}\country{France}}
\author{François Bodin}
\affiliation{\institution{%
		IRISA}\city{%
		Rennes}\postcode{35000}\country{France}}
\date{\today}

\begin{document}

\begin{abstract}
We present the mathematical foundations of a \emph{Cognitive Continuum Digital Shadow} (CCDS), a decision-support layer between users and the cross-facility infrastructure---instruments, networks, data stores and compute centers---of exascale and post-exascale scientific workflows. The CCDS couples a state-space representation of the continuum with multistage stochastic programming, so that deployment scenarios can be explored and optimized \emph{before} jobs are launched. This allows operators and users to quantify the cost, makespan and energy trade-offs of a workflow under uncertain resource availability, and hedge their decisions accordingly. We formulate the underlying optimization as a multimode, resource-constrained, stochastic supply-chain network design problem and demonstrate it on a realistic genomics workflow scheduled across heterogeneous HPC and data-center resources. This is the first of three papers; the second treats the underlying software architecture and the third reports large-scale use-cases.
\end{abstract}

\maketitle

\section{Introduction}

A digital twin (DT) is a numerical model that exchanges two-way data with its real-world counterpart \cite{asch2022DT}. Digital twins of \emph{cyberinfrastructure} itself remain rare; we are aware of the ORNL ExaDigiT platform\footnote{\url{https://exadigit.github.io/}}, the White Label Digital Twin library\footnote{\url{https://wldt.github.io/}}, and the survey \cite{DCDT2024}. With instruments such as SKA and HL-LHC coming online and their reliance on increasingly onerous cross-facility workflows, the moment is ripe to address a DT of the so-called \emph{digital continuum} \cite{asch2018pathways}: the complete, end-to-end path along which data emanates from a large instrument, is pre-processed at the edge, transmitted over a network, stored in a data center, processed in an HPC center, and results (products) are finally returned to the scientist for analysis and archiving.

Such a workflow is, by construction, executed on a cross-facility infrastructure \cite{antypas_enabling_2021,enders_superfacility_2020}---with data locality a persistent, defining constraint \cite{Unat2025}--- whose components are subject to uncertainty in availability, maintenance and cost, and to strict cybersecurity access conditions. Sustainability and the energy consumption of cyberinfrastructure \cite{IEA2025} add a further, increasingly critical constraint. A digital twin---or shadow---of the continuum (CDS) therefore offers a testbed where scheduling choices, i.e. which facilities to use under which conditions, can be explored \emph{before} the jobs are run.

This requires three software components: one to model and simulate a workflow's execution, one to schedule its sub-jobs, and one to optimize---possibly multi-objective and stochastic. The first rests on the classical state-space representation \cite{luenberger1979,friedland1985}; the second on established scheduling theory from industrial engineering \cite{brucker2007,Pinedo2022} and its stochastic extension \cite{powell2022}; the third on standard stochastic-programming theory and algorithms \cite{birgelouveaux2011,shapiro_lectures_2009}. A complete, executable treatment of the scheduling and optimization material used here, with \texttt{pyomo} implementations of every example, is provided in the open companion guide \cite{Asch2026}.

This paper is organized as follows. In Section \ref{sec:CDT} we define the concept of a digital twin, and then describe the state-space model used for the digital continuum. Then, in Section \ref{sec:SDP}, we present a detailed formulation of the stochastic, multi-stage optimization problem applied to a generic continuum setting. Section \ref{sec:sims} presents examples of stochastic multi-stage workflow optimization, applied to an academic problem and then to a realistic genomics workflow. Finally, we draw conclusions and discuss perspectives in Section \ref{sec:Conc}.

\section{Mathematical Model and Software Framework for the Continuum Digital Twin}\label{sec:CDT}

In this section, we begin by defining the twin, or shadow, that represents the digital continuum. Then we provide details of a state-space model and an optimal scheduling framework for deployment of the twin.

The notion of a digital twin has been refined, and should rather be declined at three levels: models, shadows and twins.  Models are a fixed representation, shadows are dynamic, and twins are bi-directional. A \emph{digital model} is a static, virtual representation of a physical object, system, or process. A \emph{digital shadow} is an evolving digital representation that mirrors the current state and
behaviour of the physical entity or system. It collects data from the asset through sensors, Internet of Things (IoT) devices, or
other sources that provide a feed of information that is fed into the model. This means that a digital shadow is up-to-date with the physical entity. Finally, \emph{digital  twins} integrate the virtual and physical realms by creating a real-time connection
between the physical entity and its digital counterpart, where the physical object gives
information to the digital replica and vice versa. Digital twins can then simulate, monitor, and control
physical objects or systems, facilitating analysis, optimisation, and predictive maintenance. They enable live feedback loops and foster insights for improving performance, efficiency, and reliability.

We note that in the context of the digital continuum, given cybersecurity considerations, it is highly unlikely that a digital twin could actually control the system. For this reason, in the sequel, we will concentrate on a digital shadow representation, though the term "twin" is employed.

\subsection{Mathematical Formulation: State-Space Model}

Suppose that our overarching continuum twin, the CDT, is composed
of $n$ interacting twins (or shadows)
\[
\mathrm{CDT}=\bigcup^{n}_{i=1}T_{i}
\]
 where each $T_{i}$ is itself an interacting system of $m_{i}$ sub-twins
\begin{equation}
T_{i}=\bigcup^{m_{i}}_{j=1}\tau_{ij},\quad i=1,\ldots,n.\label{eq:Twin}
\end{equation}

Examples of twins $T_{i}$ are HPC centers, data centers, data communication networks, large measurement and observation instruments (satellites, telescopes, synchrotrons, etc.) Each twin will have its particular set of sub-twins, depending on
the context and the available data. 

The state of the CDT, containing values of all pertinent state variables,
is described by a concatenated  state vector 
\[
X=\left[X_{1}\;X_{2}\;\ldots\;X_{n}\right]^{\mathrm{T}},
\]
where each component vector, $X_{i}\in\mathbb{R}^{n_{i}},$ corresponds to the
state of the $i$-th twin $T_{i}.$ The state of each twin will evolve
from time (or stage) $k$ to $k+1$ following the classical state-space
system formulation,
\begin{align}
X_{k+1} & =A_{k}X_{k}+B_{k}u_{k}+w_{k},\label{eq:SS}\\
Y_{k} & =C_{k}X_{k}+v_{k},\nonumber 
\end{align}
where
\begin{itemize}
\item $A_{k}\in\mathbb{R}^{n_{i}\times n_{i}}$ is the state transition
matrix describing the system ``dynamics'' as it evolves from time
$k$ to time $k+1.$ This matrix could be constant, even the identity
(in which case, $X_{k+1}=X_{k}$), could be an approximation of the
dynamics, could be a linearization of more complex nonlinear dynamics,
or possibly a machine-learning approximation of the state, eg. a neural
network.
\item $B_{k}\in\mathbb{R}^{n_{i}\times q_{i}}$ is the control matrix and
$u_{k}\in\mathbb{R}^{q_{i}}$ the control vector. In a classical feedback
loop, we would have $u_{k}=-KX_{k}.$ Note that in the absence of
any control, we simply set $B_{k}=0.$
\item $Y_{k}\in\mathbb{R}^{p_{i}}$ is the observed/measured state and $C_{k}\in\mathbb{R}^{p_{i}\times n_{i}}$
selects the observable state variables.
\item Finally, $w_{k}\in\mathbb{R}^{n_{i}}$ and $v_{k}\in\mathbb{R}^{n_{i}}$
represent (optional) noise, or error terms in the state and observations
respectively. The presence of these terms enables a much more detailed
quantification of the system uncertainty and the subsequent risk analysis.
\end{itemize}

\subsection{Fully coupled representation}

In reality, we need to express the coupling between components and sub-components.
At the level of the individual twin (\ref{eq:Twin}), the interaction
between its sub-components $\tau_{ij}$ can be accounted for through
an interaction matrix, or by suitable modification of the state matrix
$A_{k}$ in (\ref{eq:SS}). A new formulation can be obtained for
the coupled system by simply concatenating the individual sub-system
dynamics. 

Let us consider a given system (twin), $T,$ with $m$ sub-systems,
$\tau_{1},\ldots,\tau_{m},$ and we have dropped the index $i$ for
simplicity. Each sub-system $j$ is governed by its own state and
observations equations, denoted here by the corresponding lower-case
letters,

\begin{align}
x^{j}_{k+1} & =A^{j}_{k}x^{j}_{k}+B^{j}_{k}U^{j}_{k}+\sum_{\ell=1,\ldots,m,\,\ell\neq j}D^{j}_{\ell}x^{\ell}_{k}+w^{j}_{k},\label{eq:SS-1}\\
y^{j}_{k} & =C^{j}_{k}x^{j}_{k}+v^{j}_{k},\nonumber 
\end{align}
where $j=1,\ldots,m,$ the state $x^{j}\in\mathbb{R}^{n_{j}}$ and
we suppose that the (linear) coupling between sub-systems $p$ and
$q$ is expressed by the binary-valued term $d_{pq}$ in the coupling matrix $D^{p}_{q}\in\mathbb{R}^{n_{p}\times n_{q}},$
where $n_{i}$ is the state dimension of the sub-system $i.$ Then,
for the overall coupled system we concatenate the sub-system states
into a combined state vector $X.$ If coupling is also required at the global level of the CDT, between
the $T_{i},$ then a suitably extended form of (\ref{eq:SS-1}) can
be used instead of the system (\ref{eq:SS}). Full details are provided in \cite{MGG2025implementation}.

This representation facilitates straightforward implementation in a computer simulator, development of control algorithms, system analysis (stability, controllability, observability), state estimation techniques like Kalman filtering, multi-input, multi-output (MIMO) system analysis, and general exploration of diverse operational scenarios.

\subsection{Scheduling problem: a resource-constrained supply chain network design}

There are two ways to view the Cognitive Continuum Digital Shadow: a facility- or machine-centered view, and a job-centered view. The first is best modelled by a \emph{supply chain} model.  The second is principally a \emph{scheduling} problem.

Scheduling allocates limited resources to competing tasks over time. We adopt the standard definition \cite{Pinedo2022}.

\begin{defn}
\emph{Scheduling} is a decision-making process that deals with the
allocation of (limited) resources to tasks over given time periods.
Its goal is to optimize one or more objectives. A \emph{schedule}
is a job sequence determined for every machine of the processing system.
\end{defn}

A supply chain analog provides a model for the CCDS.

\begin{defn}
    A \emph{supply chain} is a network of suppliers, manufacturing plants, warehouses, and distribution channels organized to acquire raw materials, convert these raw materials to finished products, and distribute these products to customers.  In the CCDS context, we can consider: user-jobs, facilities, data storage, networks, job-processing and job-product delivery to end-users, respectively.
\end{defn}

\paragraph{Problem definition}
We propose a combined model for the CCDS, composed of a Supply Chain Network Design (SCND) coupled with a Resource Constrained Process Scheduling (RCPSP) model. In such a fused system, the supply chain model allocates jobs to the best facilities, and the scheduling model then (optimally) sequences jobs among facilities and within a given facility. Supply chain optimization emphasizes "where" (facility choice, data transfer mode). Scheduling emphasizes "when" and "in what order."

The resource-constrained project scheduling problem is a classical, well-known problem in operations research, and started with  the CPM (Critical Path Method) that was developed in the 1950's. A number of activities are to be scheduled. Each activity has a duration and cannot be interrupted. There are a set of precedence relations between pairs of activities which state that the second activity must start after the first has finished. The set of precedence relations are usually defined by a directed acyclic graph (DAG), where the edge $(i,j)$ represents a precedence relation where job $i$ must finish before job $j$ begins. The DAG contains two additional dummy activities with duration 0, the source and sink. Each non-renewable resource has a capacity for the entire schedule. An example would be a financial, or human resources budget that applies to the entire project. 

The multi-mode resource-constrained project scheduling problem (MRCPSP) is an extension of the resource-constrained project scheduling problem, where each task can be executed in one of a number of alternative modes. The problem aims to select a single mode from a set of available modes in order to construct a precedence- and resource-feasible project schedule with a minimal makespan. This is the similarity to a supply chain. 

\paragraph{State variables and Objective function}
The state of the system at a given time $t,$ is represented by a set of binary-valued state variables that designate the facilities that are used, at each stage, by each job to be performed in the workflow. The overall objective is then to combine location decisions---which centers to use---with allocation decisions---how to distribute the workloads and jobs among the chosen centers (data and compute). Various project performance metrics can be optimized, including task-based, resource-based, financial-based and user-based metrics. One can minimize makespan, tardiness, resource consumption, or maximize total NPV with respect to environmental sustainability, for example. 

There can be a single objective, such as minimizing (any function of) the total of fixed and variable costs, or minimizing the completion time (makespan). Multiple objective optimization (MOO) seeks a trade-off between minimum cost and  maximum sustainability (minimum environmental impact). Finally, stochastic optimization takes into account the uncertainties of resource availabilities and delays, maintenance and failures, resource allocations, variable energy costs, project costs (HR, budget).

The overall, deterministic cost function can be expressed either as a bottleneck objective, where we seek to minimize the longest or most expensive job, or as a weighted sum over all jobs of makespan, cost and any functions of these. Some terms can be ignored by setting the coefficients to zero, depending on the context.

\paragraph{RCPSP}
The Resource-Constrained Project Scheduling Problem (RCPSP) is an NP-hard combinatorial
optimization problem that consists of finding a feasible scheduling for a set of $n$ jobs subject to resource and precedence constraints.
Each job has a processing time, a set of successor jobs and a required amount of different resources. Resources may be scarce but are renewable at each time period. Precedence constraints between jobs mean that no jobs may start before all its predecessors are completed. The jobs must be scheduled non-preemptively, i.e., once started, their processing cannot be interrupted.

The RCPSP has the following input data:

\begin{itemize}
	\item  $\mathcal{J} = \{J_1, J_2, \ldots, J_n \}$ set of jobs.
	\item  $\mathcal{R}$ set of renewable resources. 
	\item   $\mathcal{S}$ set of precedences. These can be rigorously defined using an order relation, $i \prec j,$ between jobs $(i,j)\in\mathcal{J}\times\mathcal{J}.$
	\item  $\mathcal{T}$ planning horizon: set of possible processing times for jobs.
	\item  $p_{j}$ processing time of job $j.$
	\item   $u_{jr}$ amount of (renewable) resource $r$ required for processing job $j.$
	\item   $c_{r}$ capacity of renewable resource $r.$
\end{itemize}

There are many different Mixed Integer Linear Programming (MILP) formulations for the RCPSP. We choose the most suitable, discrete-time formulation. This is a binary formulation, where we have a binary variable $x_{i,t}$ for each activity $i$ and starting time/date $t,$  
$$ x_{i,t} = \begin{cases} 1, \quad \text{if activity $i$ starts on day/time $t,$} \\ 0, \quad  \text{otherwise.}  \end{cases} $$ 

The binary programming formulation, proposed by Pritsker et al. in 1986 can be written as follows.

\begin{align}
	\text{Minimize} \quad &  \sum_{t\in \mathcal{T}} t\cdot x_{n+1,t} & \label{l1}\\
	\text{Subject to:} \quad & 
	\sum_{t\in \mathcal{T}} x_{j,t}  = 1  \,\,\, \forall j\in J &  \label{l2}\\
	& \sum_{j\in J} \sum_{t_2=t-p_{j}+1}^{t} u_{jr}x_{j,t_2}  \leq c_{r}  \,\,\, \forall t\in \mathcal{T}, r \in R &  \label{l3} \\
	& \sum_{t\in \mathcal{T}} t\cdot x_{s,t} - \sum_{t \in \mathcal{T}} t\cdot x_{j,t}  \geq p_{j}  \,\,\, \forall (j,s) \in S &  \label{l4}\\
	& x_{j,t}  \in \{0,1\} \,\,\, \forall j\in J, t \in \mathcal{T}. & \nonumber
\end{align}
The objective function (\ref{l1}) represents the sum of all possible start dates for the final sink job (dummy variable) $x_{n+1,t}.$ We know that only one of these variables will equal $1$ for a specific $t.$ Therefore, by minimizing the sum of the product $t \cdot x_{n+1,t},$ we are effectively minimizing the total project duration. Constraint (\ref{l2}) ensures that each activity $i$ has exactly one start date, i.e. a single execution. Constraint (\ref{l3}) guarantees that for any time period $t,$ the schedule does not exceed the capacity $c_r$ for any renewable resource $r.$ Non-renewable resources will be introduced below in the multi-mode formulation.  Finally, constraint (\ref{l4})  ensures that if an activity $j$ follows another activity $i,$ then activity $j$ must start after the finish time of activity $i,$ which is equal to the start time of activity $i$ plus its duration $p_i.$ This is the precedence constraint.

\paragraph{MRCPSP}
The above formulation is restricted to temporal scheduling on a single machine, and for a single project, or a collection of jobs. This formulation can be generalized to deal with multiple machines and multiple projects. In this case, the formulation resembles that of a supply chain. It is referred to in the literature as the  multi-mode resource-constrained multi-project scheduling problem or MRCMPSP. 

The single-mode RCPSP assumed that each activity has only one way to be executed, whereas the multi-mode RCPSP considers multiple ways to execute an activity, which often have tradeoffs in duration, cost, or resource requirement.

Let the decision variable $x_{jm,t} \in \{0,1 \}$ denote the execution of job $j$ in mode $m$ completed in period $t.$ Then the  MRCPSP with tight, time-indexed formulation can be written as

\begin{align}
	\min \quad &  \sum_{t=\mathrm{EF}_J}^{\mathrm{LF}_J} t\cdot x_{J1,t} & \label{ll1}\\
	\text{s.t.} \quad & 
	\sum_{m=1}^{M_j}  \sum_{t=\mathrm{EF}_J}^{\mathrm{LF}_J} x_{jm,t}  = 1,  \,\,\, j=1,\ldots, J ,&  \label{ll2} \\
	& \sum_{m=1}^{M_j} \sum_{t=\mathrm{EF}_h}^{\mathrm{LF}_h} t \cdot x_{hm,t} \le \sum_{m=1}^{M_j}\sum_{t=\mathrm{EF}_j}^{\mathrm{LF}_j} (t - p_{jm}) \cdot x_{jm,t},  \,\,\, j=2,\ldots, J, \, h \in \mathcal{P}_j , &  \label{ll3} \\
	& \sum_{j=2}^{J-1} \sum_{m=1}^{M_j} k_{jmr}^{\rho} \sum_{q=\max\{t,\mathrm{EF}_j\} }^{\min \{ t+p_{jm}-1,\mathrm{LF}_j \}}  x_{jm,q} \le K_r^{\rho}, \,\,\,  r\in R^{\rho}, \, t=1,\ldots, \bar{T}, &  \label{ll4} \\
	& \sum_{j=2}^{J-1} \sum_{m=1}^{M_j} k_{jmr}^{\nu}   \sum_{t=\mathrm{EF}_J}^{\mathrm{LF}_J}  x_{jm,t} \le K_r^{\nu}, \,\,\, r\in R^{\nu},  &  \label{ll5}\\
	& x_{jm,t}  \in \{0,1\}, \,\,\, j=1,\ldots,J, \, m=1,\ldots ,M_j, \, t =0,\ldots, \bar{T}, &  \nonumber
\end{align}
where $\mathcal{P}_j$ is the set of predecessors of job $j,$  the earliest finish and latest finish times of job $j$ are denoted $\mathrm{EF}_j,$ $\mathrm{LF}_j,$ an upper bound on the project's makespan is given by $\bar{T}$, and we have denoted renewable resources by the index $\rho$ and non-renewables by $\nu.$

Note that objective (\ref{ll1}) is the minimization of the makespan, to which we can add any cost function of the duration, such as energy consumption. The constraints  (\ref{ll2}) indicate that each activity is assigned exactly one mode and exactly one finish time;  (\ref{ll3}) ensures that no activity is started until all its predecessors are finished;  (\ref{ll4}) ensures that the per-period levels of the renewable resources are met;   consumption of the nonrenewable resources is limited to their availabilities by  (\ref{ll5}). Finally, restricting the summation produces what is known as a tighter formulation over time to the intervals $[\mathrm{EF}_j, \mathrm{LF}_j]$ that reduces drastically the number of variables and gives better convergence. These time windows can be computed by simple forward and backward recursion loops.

To include infrastructure costs in the objective function, we just add a term, or terms, if other costs are to be taken into account, e.g. related to sustainability,
\begin{align}
	\min \quad &  \sum_{t=\mathrm{EF}_J}^{\mathrm{LF}_J} t\cdot x_{J1,t} 
	+ \sum_{j,m} c_{jm} x_{jm} , \nonumber
\end{align}
where $c_{jm}$ is the cost associated with the use of facility $m$ for task $j.$

We can also perform simultaneous scheduling of a set of multiple projects taking into account the availability of local and global resources under different time and resource constraints. This has practical importance, at national and European levels, when cross-facility implies exploitation of cyberinfrastructure resources across different countries, for example, as would be the case for EuroHPC  \footnote{\url{https://www.eurohpc-ju.europa.eu}} at the European level, or GENCI \footnote{\url{https://www.genci.fr/}} for France. 

This general formulation of the RCPSP is then mathematically equivalent to the supply chain configuration problem SCCP with the addition of resource constraints. The loop is closed. This formulation is the basis for the multistage stochastic model of Section \ref{sec:SDP} below. 

\paragraph{Sensitivity analysis}
Duality theory in linear-programming makes the model interpretable \cite{Boyd2004,Nocedal2006}. By strong duality, the optimal Lagrange multiplier $\lambda_j^*$ associated with a binding constraint satisfies the sensitivity identity
\[
\lambda_j^* = \frac{\partial f^*}{\partial b_j},
\]
where $f^*$ is the optimal objective value and $b_j$ the right-hand side of constraint $j.$ Applied to the capacity constraints (\ref{ll4}-\ref{ll5}), these multipliers are \emph{shadow prices}: the marginal reduction in total cost obtainable from one additional unit of capacity at facility $m$, with $\lambda_m^*=0$ signaling slack, reallocatable capacity. These sensitivities give facility operators and funding agencies a principled, quantitative basis for capacity-planning decisions; the duality and KKT background, with worked \texttt{pyomo} examples, is given in \cite{Asch2026}.

\subsection{Three-Module Software Framework}

\begin{figure}
\begin{centering}
\includegraphics[width=0.75\textwidth]{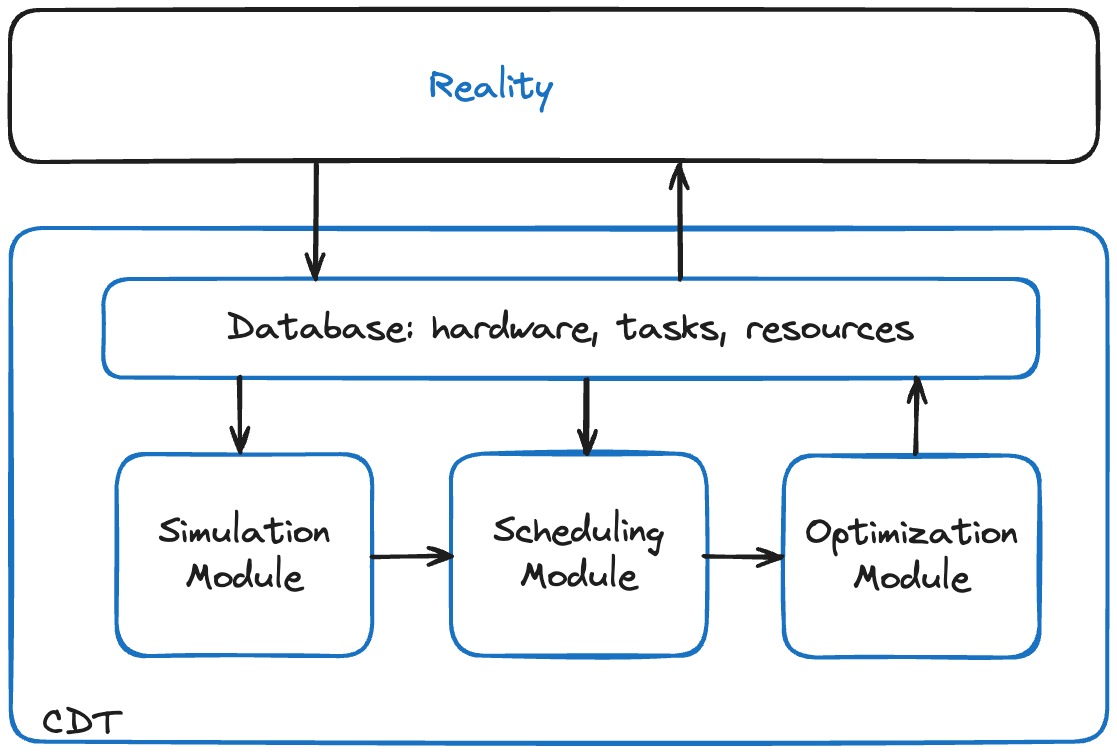}
\par\end{centering}
\caption{Global software architecture of the CDT, consisting of three modules:
simulation (SIM), scheduling (SCD), optimization (OPT).\label{fig:Global-software-architecture}}
\end{figure}

The global software framework for the CDT is composed of a set of
three modules: simulation (SIM), scheduling (SCD) and optimization (OPT), as shown in Figure \ref{fig:Global-software-architecture}. 
They can be used independently, or be chained together. These three are fed by
a shared database, based on a common ontology that contains descriptions of all the jobs to perform,
resources available, constraints to be respected, and any objectives
to be attained, which together constitute the \emph{cognitive} part. This database is connected to, and communicates with
the real world---see Figure \ref{fig:Global-software-architecture}.
This communication can take the form of a MADPP (machine actionable
data project plan), a user-interface, or a combination of the two \cite{MGG2025implementation}.

\begin{description}
    \item[SIM] The simulation module implements the state-space model (\ref{eq:SS}). It is fully detailed in \cite{MGG2025implementation}.
    \item[SCD] The scheduling module (SCD) computes optimal scheduling models, adapted to the infrastructure and the choice of objectives. This module proposes a stochastic approach.
    \item[OPT] The optimization module (OPT) serves as a basis for the scheduling module, and can also be used independently. For example, the user may prefer a heuristic approach, such as first in first out (FIFO), that does not require any scheduling as such. Multi-objective optimization (MOO) can also be performed directly, without scheduling. In the most general case, a stochastic optimization based on a multistage formulation is implemented---see Section \ref{sec:SDP}.
\end{description}

We envisage two roles for the OPT and SCD  modules in the CDT:
\begin{enumerate}
    \item A strategic, offline role.
    \item A tactical, online role.
\end{enumerate}
The first is to be used for planning a very large scale workflow, \emph{before} its execution. The second will serve as a real time optimizer or scheduler to be used in conjunction with the SIM module. As a result, different approaches will be used for each role: stochastic, multi-stage programming for the former, simpler heuristics\footnote{Commonly used by batch schedulers in HPC centers.} for the latter, such as FIFO, fair share, backfill, etc. The former will be illustrated below in the examples Section \ref{sec:sims}, and the latter in \cite{MGG2025implementation,MGG2025usecases}.

\paragraph{Practical realization.}
The SCD/OPT pipeline is realized in four steps: (i) a generic \textsc{json} description of the problem instance---resources with renewable and non-renewable capacities, jobs with release times and deadlines, alternative execution \emph{modes} per job (the candidate facilities, with their durations, costs and resource requirements), and the precedence DAG---validated against a published \textsc{json} schema; (ii) a model-generator function that translates this description into (iii) a \texttt{pyomo} \cite{bynum2021pyomo} optimization model; whose solution yields (iv) the optimal schedule, exported as \textsc{csv}/\textsc{json}. The deterministic scheduling kernel is a multi-mode resource-constrained project scheduling problem (MRCPSP) \cite{brucker2007,Pinedo2022}, with machine-conflict disjunctions handled by big-M or generalized disjunctive programming reformulations. The complete data model, schema, code and worked examples---from economic dispatch through two-stage and multistage stochastic programs to risk-averse dispatch---are available in the open, executable companion guide \cite{Asch2026}. Programmatic access to the facilities themselves can rely on existing RESTful interfaces to HPC systems such as FirecREST \cite{cruz_firecrest_2020}.

\section{Stochastic, multistage programming}\label{sec:SDP}

\subsection{Background}
Stochastic programming addresses decision-making under uncertainty: some decisions must be taken now, before the information that would justify them becomes available \cite{powell2022,shapiro_lectures_2009}. Its central notion is the \emph{stage}---a point in time at which decisions are made, separating what is known from what is not. A \emph{two-stage} problem makes one major decision now and adapts the rest once uncertainty is revealed; a \emph{multistage} problem repeats this here-and-now/wait-and-see pattern over several stages. In the CCDS, the first stage fixes the major facility commitments and later stages adapt job routing as availability is observed.

\subsection{Mathematical Formulation}

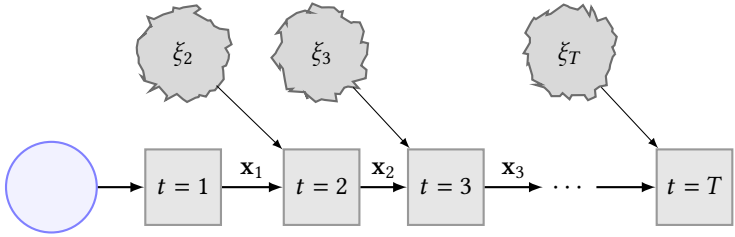
\begin{figure}[htbp]
	\centering
    \tikzstyle{state}=[circle,
		thick,
		minimum size=1.2cm,
		draw=blue!50,
		fill=blue!5]	
\tikzstyle{matrx}=[rectangle,
		thick,
		minimum size=1.0cm,
		draw=gray!80,
		fill=gray!20]	
\tikzstyle{noise}=[circle,
		thick,
		minimum size=1.2cm,
		draw=gray!85!black,
		fill=gray!30,
		decorate,
		decoration={random steps,
			segment length=2pt,
			amplitude=2pt}]
		
\begin{tikzpicture}[>=latex,text height=1.5ex,text depth=0.25ex]
			\matrix[row sep=0.5cm,column sep=0.5cm] {
				&
				\node (xi_2) [noise] {$\mathbf{\xi}_{2}$};    &
				\node (xi_3)   [noise] {$\mathbf{\xi}_3$};     &
				&
				\node (xi_T) [noise] {$\mathbf{\xi}_{T}$};   
				\\
				\node (t0)  [state]   {$ $};           &
				\node (t_1) [matrx] {$t=1$};       &
				\node (t_2) [matrx] {$t=2$};      &
				\node (t_3) [matrx] {$t=3$};      &
				\node (t_4)         {$\cdots$};       &
				\node (t_T) [matrx] {$t=T$};       \\
			};			
			\path[->]
			(t0) edge[thick] (t_1)	
			(t_1) edge[thick] node  [above]{$\mathbf{x}_1$} (t_2)	 
			(t_2) edge[thick] node  [above]{$\mathbf{x}_2$} (t_3)		
			(t_3)   edge[thick] node  [above]{$\mathbf{x}_3$} (t_4)		
			(t_4)   edge[thick] (t_T)	
			(xi_2) edge (t_2)				
			(xi_3) edge (t_3)
			(xi_T) edge (t_T)
			;			
\end{tikzpicture}	
	\caption{A multistage decision process with uncertainty. Uncertain, exogenous inputs $\xi_t$ arrive, starting from stage $t=2,$ and trigger subsequent recourse actions.}
    \label{fig:SDP}
\end{figure}

In Figure \ref{fig:SDP}, we depict a generic multistage  decision process with state $\mathbf{x}_t$ that evolves over time, starting at stage $t=1,$ and then receives uncertain, exogenous inputs, $\xi_2, \xi_3, \ldots, \xi_T,$ at given stages $ t = 2,3,\ldots, T.$ At stage 1, a here-and-now (deterministic) decision is taken. The subsequent sequence of wait-and-see decision functions ${\mathbf{x}_t(\xi_t)}_{t\in [T]} $ constitutes a \emph{policy}, $\pi.$ The policy thus obtained, provides a decision rule for all stages $ t \in [T].$ The aim of the decision process (stochastic multistage optimization) is then to compute an \emph{optimal} policy subject to a given objective and constraints.

We will describe in detail the two-stage problem, since the multistage is a straightforward generalisation. 
In the stochastic setting, we can naturally classify $x\in\mathscr{X}$ as the \emph{first-stage} decision
variables, since the major centres are fixed, and $y\in\mathbb{R}^{\left|\mathscr{A}\right|\times\left|\mathscr{K}\right|}$
as the \emph{second-stage} decision variables, since the job-flow operating conditions are uncertain.

The two-stage stochastic linear program \cite{birgelouveaux2011, shapiro_lectures_2009} is then
\begin{align} \label{eq:2stage}
\min_{x\in\mathscr{X}} & \,\,c^{\top}x + \mathrm{E}\left[Q(x;\xi)\right],
\end{align}
where $Q(x;\xi)$ is the optimal value of the second-stage problem
\begin{align*}
\min_{y\ge0} & \,\,q^{\top}y\\
\textrm{s.t.} & \,\,Ny=0,\\
 & Cy\ge d,\\
 & Sy\le s,\\
 & Ry\le Mx,
\end{align*}
where $d$ is the (uncertain) demand, $s$ is the (uncertain) supply, the random vector $\xi=(q,d,s,R,M)$ and $y=y(\xi).$ The expectation in (\ref{eq:2stage}) is taken with respect to the joint probability distribution function of $\xi.$  In the terminology of state-space, as used in Section \ref{eq:SS}, we can consider $x$ as the \emph{state} variable, and $y$  as the \emph{control} variable.

To treat the occurrence of infeasibility, where the first-stage solution
does not satisfy the second-stage constraints, e.g. $Cy\nleqslant d,$
we can introduce a \emph{recourse} action $z$  that supplies
the deficit $d-Cy$ at some penalty cost. Then the second-stage problem becomes
\begin{align*}
\min_{y\ge0} & \,\,q^{\top}y+h^{\top}z\\
\textrm{s.t.} & \,\,Ny=0,\\
 & Cy+z\ge d,\\
 & Sy\le s,\\
 & Ry\le Mx,
\end{align*}
where $h$ represents the vector of (positive) recourse costs.

\begin{rem}
     Scheduling---via the module SCD---provides the cost function, then stochastic linear programming---via the module OPT---performs the minimization.    
\end{rem}

In the multistage case, we have a succession of recourse stages, yielding the multistage program \cite{powell2022}, 

\[
      \min_{\pi \in \Pi} \mathrm{E} \left[ \sum^{T}_{t=1} \gamma^{t-1} f_t(x_t;\xi_t) \right],
\]
subject to
\[
     x_t \in \mathcal{X}_t(x_{t-1};\xi_{t-1}), \quad t=1 \ldots, T,
\]
where $\pi$ is a policy, $\gamma$ a discount factor, $f_t$ a stage cost, $x_t$ the stage decision and $\mathcal{X}_t$ its constraint set. A policy is a decision rule that optimizes the objective over all stages.

The sense in which the objective is minimized must be specified. The weakest is the \emph{expectation} sense, e.g. minimizing $\mathrm{E}[f(C_{\max})]$; a stronger \emph{stochastic} sense requires $f(C_{\max})$ to be smaller in distribution than under any other policy \cite{Pinedo2022}.

\subsection{Beyond expectation: chance constraints, risk and robustness}\label{sec:risk}

Minimizing an expectation is risk-neutral. Cross-facility deployments often require stronger guarantees. Three standard extensions, all expressible within the OPT module, are treated in detail---with runnable \texttt{pyomo} implementations---in the companion guide \cite{Asch2026}.

\emph{Chance constraints} require uncertain constraints to hold with prescribed probability,
\[
\mathrm{P}\left( a_i(\xi)^{\top}x - b_i(\xi) \le 0,\ i=1,\ldots,m \right) \ge 1-\epsilon,
\]
for a tolerance $\epsilon\in(0,1)$---for example, guaranteeing that a deadline is met in at least $95\%$ of availability scenarios.

\emph{Risk measures} replace the expectation by a functional $\rho$ that penalizes dispersion or tail losses: the mean--variance criterion $\mathrm{E}(Z)+\kappa\,\mathrm{Var}(Z)$ with risk-aversion parameter $\kappa$, or the conditional value-at-risk
\[
\mathrm{CVaR}_{\alpha}(Z)=\mathrm{E}\left[\,Z \mid Z \ge \mathrm{VaR}_{\alpha}(Z)\,\right],
\qquad
\mathrm{VaR}_{\alpha}(Z)=\inf\{t : \mathrm{P}(Z\le t)\ge\alpha\},
\]
which controls the expected cost in the worst $(1-\alpha)$ fraction of scenarios \cite{shapiro_lectures_2009}. CVaR admits a linear-programming reformulation and thus integrates directly into the supply-chain model of Section~\ref{sec:CDT}.

\emph{Robust optimization} dispenses with distributions altogether and optimizes against the worst case over an uncertainty set; it yields stabler but, on average, more conservative schedules. The choice among these formulations is a modeling decision exposed to the CCDS user, who can trade average performance against protection in the tail---precisely the trade-off faced when an urgent-computing deadline coexists with energy budgets.

\section{Examples of workflow optimization}\label{sec:sims}

In this section we present two examples. The first is a well-known academic example that expresses, in a simple way, the basic principles and interest of  stochastic optimization. The second example represents a realistic case of a cross-facility genomics workflow.

\subsection{Newsvendor problem as a prototype}

We choose this problem because of its simplicity, the fact that it can be solved analytically, and its representativity of a wide diversity of stochastic optimization problems. In particular, those that arise in the context of the CCDS.

The simplest stochastic multistage decision problem is the two-stage setting defined in (\ref{eq:2stage}). The newsvendor has to decide, at stage 1, how many newspapers to purchase. Then, at the later stage 2, the stochastic demand is revealed that will determine his overall profit or loss. The vendor wants to take a decision that will \emph{hedge} their exposure to the risk of under- or overstocking, while taking into account the uncertain demand.

It can readily be shown that the optimal order quantity, $x_*,$ is given by
\[
   x_* = F^{-1} ( \mathrm{CR} ), 
\]
where the critical (cost) ratio (CR) is defined as
\[
  \mathrm{CR} = \frac{C_{\mathrm{under}} }{C_{\mathrm{under}} + C_{\mathrm{over}} },
\]
the ratio between understocking costs and overstocking costs. These costs are defined in terms of the basic pricing parameters: purchase price, selling price, salvage price, and $F$ is the cumulative probability distribution function of the randomly varying demand $\xi.$ 

\begin{figure}
\begin{centering}
\includegraphics[width=0.8\textwidth]{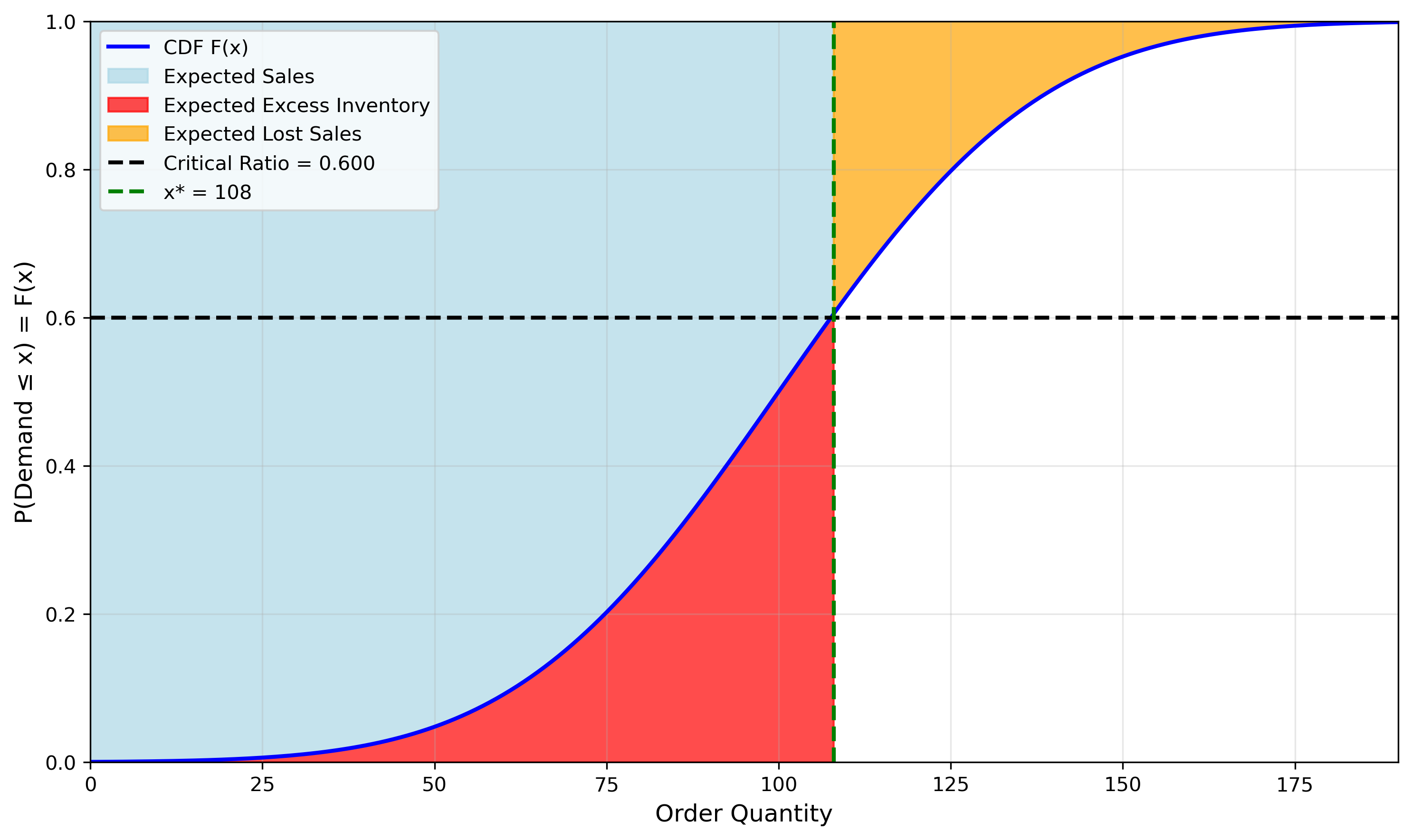}
\par\end{centering}
\caption{Simulation of 2-stage stochastic optimization for the newsvendor problem. By modifying CR and $\sigma,$ it is possible to explore different hedging positions, i.e. whether to buy more or less than the average demand $\mu $ (depends on the value of CR), and by how much (depends on magnitude of $\sigma.$) \label{fig:newsvendor}}
\end{figure}

For illustration, suppose that the demand, $\xi,$ follows a Gaussian distribution 
\[
   \xi \sim \mathcal{N} (\mu, \sigma^2),
\]
with mean $\mu$   and variance $\sigma^2,$  whose values are estimated from past sales data. Stochastic optimization results for this case are shown in Figure \ref{fig:newsvendor}, where the vendor can evaluate the trade-off by modifying their stock risk aversion, CR, and the demand variance $\sigma^2.$  This produces a stochastically optimal hedging position, as well as clearly showing the expected excess inventory and lost sales. 

This foundational model can then be readily applied to different elements within the CCDS context. For example, how many servers, or partitions, to allocate to a workflow, while taking into account uncertainty in availability, or waiting times, or any other factor that has an incidence on the global cost of the workflow, or its makespan. 

\subsection{A Genomics Workflow Use-Case}

We now consider a slightly simplified, but nonetheless realistic example of a genomics workflow \cite{Jian2015, tanjo_practical_2021}. Suppose that we have four jobs to perform: preprocessing ($J_1$), alignment ($J_2$), variant calling ($J_3$), and annotation ($J_4$). These have the following DAG structure:

\begin{center}
\includegraphics[width=0.275\paperwidth]{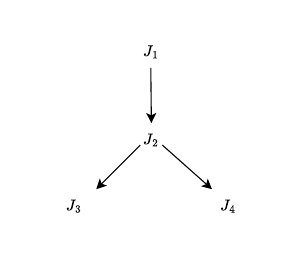}
\par\end{center}

The descriptions of the jobs, their resulting datasets and the CPU times needed for processing are summarized in Table \ref{tab:genomics}.

\begin{table}
\begin{centering}
\begin{tabular}{|c|c|c|c|c|}
\hline 
  Job & Description &  Input & Output & Processing time \tabularnewline
\hline 
\hline 
$J_{1}$ & \begin{cellvarwidth}[t]
\centering
Preprocessing raw \\ sequencing data
\end{cellvarwidth} & \begin{cellvarwidth}[t]
\centering
$D_{1}$\\ raw read\\ 100 GB
\end{cellvarwidth} & \begin{cellvarwidth}[t]
\centering
$D_{2}$\\ cleaned reads \\ 80 GB
\end{cellvarwidth} &  10h\tabularnewline
\hline 
 $J_2$  & \begin{cellvarwidth}[t]
\centering
Read \\ alignment
\end{cellvarwidth} & \begin{cellvarwidth}[t]
\centering
$D_{2}$\\ 80 GB
\end{cellvarwidth} & \begin{cellvarwidth}[t]
\centering
$D_{3}$\\ aligned BAM\\ 50 GB
\end{cellvarwidth} & 20h \tabularnewline 
\hline 
  $J_3$ & \begin{cellvarwidth}[t]
\centering
Variant \\ calling
\end{cellvarwidth} & \begin{cellvarwidth}[t]
\centering
$D_{3}$\\ 50 GB
\end{cellvarwidth} & \begin{cellvarwidth}[t]
\centering
$D_{4}$\\ VCF\\ 5 GB
\end{cellvarwidth} & 15h \tabularnewline 
\hline 
  $J_4$ & \begin{cellvarwidth}[t]
\centering
Functional \\ annotation
\end{cellvarwidth} & \begin{cellvarwidth}[t]
\centering
$D_{3}$\\ 50 GB
\end{cellvarwidth} & \begin{cellvarwidth}[t]
\centering
$D_{5}$\\ annotated\\ 10 GB
\end{cellvarwidth} & 10h \tabularnewline 
\hline 
\end{tabular}
\par\end{centering}
\caption{Specifications for jobs $J_i$ and datasets $D_j$ in a genomics workflow. All terms and abbreviations are standard \cite{Jian2015, tanjo_practical_2021}. \label{tab:genomics}}

\end{table}

The cyberinfrastructure is composed of two datacenters, say AWS and Google cloud,  and two computing infrastructures, say a local university server (cheaper, but less reliable/available) and a cloud HPC service (expensive, but high availability). These are denoted $\mathrm{DC}_1,$  $\mathrm{DC}_2,$ and $\mathrm{HP}_1,$  $\mathrm{HP}_2,$ respectively.

\newpage
When the probability distributions for the random parameters (events) are discrete, there are only a finite number of outcomes in each stage of the stochastic program. With each random parameter fixed to one of its possible outcomes, one can create a scenario representing one possible realization of the future. Enumeration of all possible combinations of outcomes allows us to represent all scenarios in a tree, with each scenario being a path from the root of the tree to one of its leaves. The nodes visited by each path correspond to values assumed by random parameters in the model.

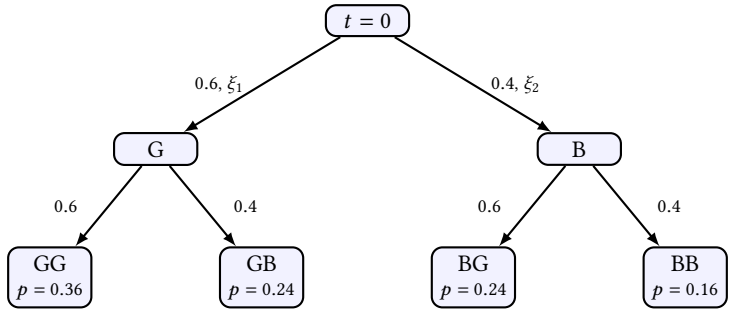
\begin{figure}[htbp]
\centering
\begin{tikzpicture}[
  grow=down, ->, >=latex, level distance=17mm,
  level 1/.style={sibling distance=56mm},
  level 2/.style={sibling distance=28mm},
  every node/.style={draw, rounded corners, thick, fill=blue!5, minimum width=11mm, inner sep=3pt, font=\small, align=center},
  edge from parent/.style={draw, thick},
  lab/.style={draw=none, fill=none, font=\scriptsize, midway}
]
\node {$t=0$}
  child { node {G} 
    child { node {GG\\[-2pt]{\scriptsize $p=0.36$}} edge from parent node[lab,left]{$0.6$} }
    child { node {GB\\[-2pt]{\scriptsize $p=0.24$}} edge from parent node[lab,right]{$0.4$} }
    edge from parent node[lab,left]{$0.6$,\ $\xi_1$}
  }
  child { node {B} 
    child { node {BG\\[-2pt]{\scriptsize $p=0.24$}} edge from parent node[lab,left]{$0.6$} }
    child { node {BB\\[-2pt]{\scriptsize $p=0.16$}} edge from parent node[lab,right]{$0.4$} }
    edge from parent node[lab,right]{$0.4$,\ $\xi_2$}
  };
\end{tikzpicture}
\caption{Scenario tree for the genomics workflow. G: all centers available, $c_{\mathrm{HP}_1}=$ 100€; B: $\mathrm{HP}_1$ unavailable, jobs route to $\mathrm{HP}_2$ at $c_{\mathrm{HP}_2}=$ 250€. Leaves define the four scenarios $S_1,\ldots,S_4$.\label{fig:scentree}}
\end{figure}

The scenario tree, shown in Figure \ref{fig:scentree}, has two branching stages: at each of $t=1$ and $t=2$, the system is either Good (G, all centers available) with probability $0.6$, or Bad (B, $\mathrm{HP}_1$ unavailable) with probability $0.4$.

The four leaf nodes define the scenarios:
\begin{itemize}
    \item $S_1$ (GG): $p=0.6 \times 0.6 = 0.36,$ all systems available, normal costs.
    \item $S_2$ (GB): $p=0.6 \times 0.4 = 0.24,$ $t=1$ good, $t=2$ $\mathrm{HP}_1$  unavailable.
    \item $S_3$ (BG): $p=0.4 \times 0.6 = 0.24$, $t=1$ bad, $t=2$ recovery.
    \item $S_4$ (BB): $p=0.4 \times 0.4 =0.16,$ persistent outages.
\end{itemize}

The operational costs---these could also be expressed in terms of energy consumption if required---for the job execution and network transfers are: \\

\hspace{1cm}
\begin{tabular}{ccc}
     Job & $\mathrm{HP}_1$ & $\mathrm{HP}_2$ \\ \hline 
     $J_1$ & \texteuro100 & \texteuro250 \\
     $J_2$ & \texteuro200 & \texteuro500 \\
     $J_3$ & \texteuro150 & \texteuro400 \\
     $J_4$ & \texteuro100 & \texteuro300 \\  
\end{tabular}
\hspace{2cm}
\begin{tabular}{cc}
     From $\rightarrow$ To & Cost \\ \hline 
     $\mathrm{DC}_1 \rightarrow \mathrm{DC}_2$  & \texteuro0.50/GB \\
     $\mathrm{DC}_2 \rightarrow \mathrm{DC}_1$ & \texteuro0.50/GB  \\
     $\mathrm{DC}_j \rightarrow \mathrm{HP}_i$ & \texteuro0.20/GB  \\
     $\mathrm{HP}_i \rightarrow \mathrm{DC}_j$ & \texteuro0.10/GB   \\  
\end{tabular} \\ \\
\noindent In addition, there are fixed storage costs, $\gamma_{ij},$ for dataset $i$ at location $j$ in \texteuro{} per GB per stage for any location, equal to \texteuro0.05/GB/day. 

We can now formulate the multistage stochastic optimization problem in compact formulation, according to the scenario tree, based on (\ref{ll1}-\ref{ll5}) and (\ref{eq:2stage}). Let $n \in \{1,2,3,4\}$ index the leaf scenarios. Then the three decision variables are:
\begin{enumerate}
    \item $x_{jh,n}^t$ denoting execution of job $j$ on HPC $h$ in stage $t$ under scenario $n,$
    \item $y_{idk,n}^t$ denoting the transfer of dataset $i$ from $d$ to $k$ in stage $t,$ 
    \item $s_{id,n}^t$ denoting the storage of dataset $i$ at location $d$ at the end of stage $t.$
\end{enumerate}
The objective function is the expectation---probability weighted sum---of the global processing costs $c_{jh},$ the global transfer costs $\tau_{dk}$ and the global (fixed coefficient) storage costs,

$$\min_{x,y,s} \sum_{n=1}^4 p_n \left[\sum_{t=1}^3 \left(\sum_{j,h} c_{jh} x_{jh,n}^t + \sum_{i,d,k} \tau_{dk} y_{idk,n}^t + \sum_{i,d} 0.05 \cdot s_{id,n}^t\right)\right],$$
subject to the following constraints for each scenario $n.$
\begin{enumerate}
    \item Job execution once:
$$\sum_{t=1}^3 \sum_{h \in \{\text{HP}_1,\text{HP}_2\}} x_{jh,n}^t = 1, \quad \forall j .$$
    \item Precedence ($J_1$ before $J_2$):
$$\sum_{\tau=1}^{t-1} \sum_h x_{J_1,h,n}^\tau \geq x_{J_2,h,n}^t, \quad \forall h, t.$$
    \item Data availability ($J_2$ needs $D_2$ at execution site):
$$s_{D_2,h,n}^{t-1} \geq 80 \cdot x_{J_2,h,n}^t, \quad \forall h, t .$$ 
    \item Data balance at $\text{DC}_1$:
$$s_{D_1,\text{DC}_1,n}^t = s_{D_1,\text{DC}_1,n}^{t-1} - \sum_k y_{D_1,\text{DC}_1,k,n}^t + \sum_k y_{D_1,k,\text{DC}_1,n}^t . $$
    \item Output generation ($J_1$ creates $D_2$):
$$s_{D_2,h,n}^t = s_{D_2,h,n}^{t-1} + 80 \cdot x_{J_1,h,n}^t + \text{[transfers]}. $$
    \item Availability constraints:
$$x_{jh,n}^t \leq \alpha_{h,n}^t,$$ where the availability $\alpha_{h,n}^t(\xi^t)$ of HPC center $h$ at stage $t$  is a binary value that expresses the availability (0 or 1) as defined in the scenario tree.
    \item Non-anticipativity (critical): at $t=1$, scenarios sharing same history must have same decisions,
$$x_{jh,S_1}^1 = x_{jh,S_2}^1, \quad x_{jh,S_3}^1 = x_{jh,S_4}^1 . $$
\end{enumerate}

We execute the program using pyomo \cite{bynum2021pyomo,hart2011pyomo} and the GLPK\footnote{\url{https://www.gnu.org/software/glpk/}} LP/MIP Solver 5.0. Execution time on a laptop, for 1888 variables and 976 constraints, is less than $0.1$ seconds. The optimal scheduling results for the above data are presented in Figure \ref{fig:genomics_results}. We observe the following (see details in the Appendix):
\begin{itemize}
    \item In Stage 1, hedging leads to the choice of the more reliable $\text{HP}_2$ initially, despite higher cost.
    \item Adaptive recourse switches to cheaper $\text{HP}_1$  when available at stage 2.
    \item Data locality is privileged by keeping intermediate data at execution sites to minimize transfers.
    \item Non-anticipativity is ensured since the solution cannot ``cheat" by using future information.
    \item Finally, we obtain a risk-return tradeoff, since the stochastic solution balances expected cost vs. robustness.
\end{itemize}

The value of this stochastic formulation is quantified by two standard diagnostics \cite{birgelouveaux2011}. Let $\mathrm{RP}$ denote the optimal value of the recourse problem above, $\mathrm{WS}$ the (clairvoyant) wait-and-see value obtained by solving each scenario with perfect foresight, and $\mathrm{EEV}$ the expected cost of naively implementing the solution of the mean-value problem. Then the \emph{expected value of perfect information}, $\mathrm{EVPI}=\mathrm{RP}-\mathrm{WS}$, bounds what one should pay for better availability forecasts---directly pricing the predictive applications of the CCDS---while the \emph{value of the stochastic solution}, $\mathrm{VSS}=\mathrm{EEV}-\mathrm{RP}$, measures the cost of ignoring uncertainty altogether. Both are computed routinely by the OPT module; implementations are provided in \cite{Asch2026}.

We can also perform a \emph{multi-objective optimization} for this workflow. This would be deterministic case, where we seek an optimal trade-off---or Pareto front---between (for example) minimum time to completion and minimum energy costs, supposing we have such data for each data center and HPC center. Another trade-off could be energy costs vs. center utilization, for example when urgent computing is requested. All of these problems can be formulated and solved by the OPT and SCD modules of the CCDS.

\begin{figure}
\begin{centering}
\includegraphics[width=1.0\textwidth]{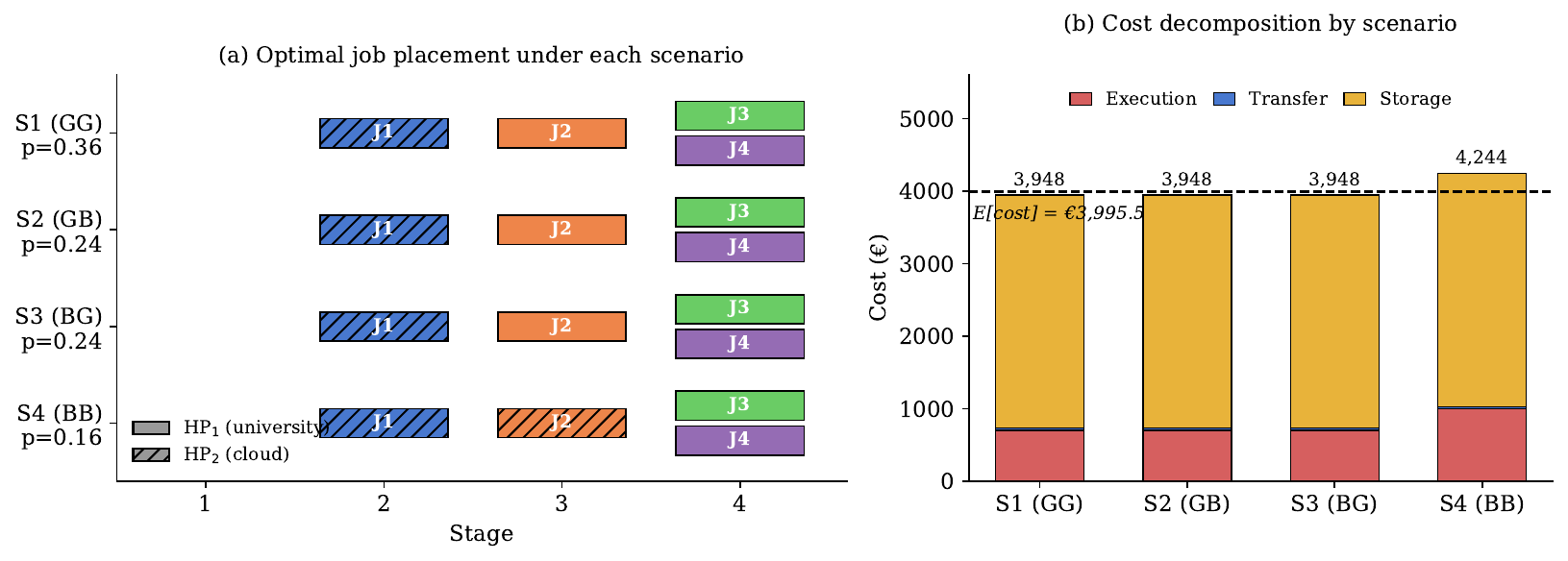}
\par\end{centering}
\caption{Multistage stochastic optimization of the genomics workflow. (a) Optimal job placement in each scenario: the reliable but costlier $\mathrm{HP}_2$ is chosen as a hedge in stage~2, with recourse to the cheaper $\mathrm{HP}_1$ once availability is observed; only the persistent-outage scenario $S_4$ retains $\mathrm{HP}_2$ for $J_2$. (b) Cost decomposition per scenario; the dashed line marks the expected total cost. Scenarios and variables are defined in the text.\label{fig:genomics_results}}
\end{figure}

Admittedly the modeling is the greatest challenge here, but starting from a simple setting---such as the one just described above---it is relatively easy to complexify the model and attain more and more realism. A structured approach for this is proposed in \cite{powell2022}.

Detailed use-case applications of the CCDS are presented in \cite{MGG2025usecases}.

\section{Conclusions and outlook}\label{sec:Conc}

A full exascale workflow cannot be tested or optimized before deployment, since access to HPC, networks and data centers is constrained by cybersecurity and resource-allocation policies, and availability is uncertain. A digital shadow of the continuum is therefore an indispensable planning tool. We have presented its mathematical foundations---a state-space model of the continuum coupled with multistage stochastic and multi-objective optimization---and a software framework realizing them, validated here on a genomics workflow and, at full scale, in the companion use-cases \cite{MGG2025usecases}.

\subsection{Impact}

By integrating simulation, scheduling and stochastic optimization into a single Continuum Digital Twin, the CCDS turns ad-hoc, expert-driven deployment choices into a reproducible, quantitative decision process. Its main contributions are robust planning under uncertain resource availability, explicit cost--makespan--energy trade-offs, and a generic System-of-Systems formulation that transfers across HPC centers, large instruments and global cyberinfrastructure.

\subsection{Outlook and perspectives}

The framework presented here is just a start. The real challenge lies in its adoption by existing (and future) large-scale scientific projects. In this optic, we are currently in contact with the following project groups:

\begin{enumerate}
    \item SKA (\url{https://www.skao.int/}).
    \item CERN-LHC (\url{https://home.cern/}).
    \item GAIA-data (\url{https://www.gaia-data.org/}).
\end{enumerate}

Prototypes of the CCDS are currently under development \cite{MGG2025usecases} for two concrete pipelines, also documented in \cite{Asch2026}: DDFacet, a wide-field radio-interferometric imaging workflow relevant to SKA-precursor data, and NSBAS, an InSAR time-series processing chain within GAIA-Data. Discussions are underway with LHC. These will lead us to work with EuroHPC (\url{https://www.eurohpc-ju.europa.eu/}) as the underlying digital infrastructure.

\section*{Acknowledgements}

As part of the France 2030 initiative,
this work has benefited from a national grant managed by the French
National Research Agency (Agence Nationale de la Recherche) attributed
to the ExaAToW project of the NumPEx PEPR program, under the reference
ANR-22-EXNU-0005. 

\section*{Software and Data Availability}

All code, together with explanations and worked examples, is available in the open, executable companion guide \emph{Optimal Scheduling for Cross-Facility Workflows} \cite{Asch2026}, \url{https://markasch.github.io/RCP4CDT/}, with sources at \url{https://github.com/markasch/RCPSP4CDT}.

\section*{AI Disclosure}
The code for the example of the genomics workflow was partially generated using Claude 3.5 Sonnet (accessed November 2025). All AI-generated code was reviewed, modified, tested, and revised by the authors, who confirm accuracy and reproducibility of results.

\bibliographystyle{acm}
\bibliography{CDT}

@ARTICLE{DCDT2024,
author={Athavale, Jyotika and Bash, Cullen and Brewer, Wesley and Maiterth, Matthias and Milojicic, Dejan and Petty, Harry and Sarkar, Soumyendu},
journal={ Computer },
title={{ Digital Twins for Data Centers }},
year={2024},
volume={57},
number={10},
ISSN={1558-0814},
pages={151-158},
doi={10.1109/MC.2024.3436945},
url = {https://doi.ieeecomputersociety.org/10.1109/MC.2024.3436945},
publisher={IEEE Computer Society},
address={Los Alamitos, CA, USA},
month=oct}

@misc{IEA2025,
    title = {Energy and AI},
    author = {IEA,  Paris},
    year = {2025},
    url={https://www.iea.org/reports/energy-and-ai}, 
}

@book{asch2022DT,
	title = {A {Toolbox} for {Digital} {Twins}: {From} {Model}-{Based} to {Data}-{Driven}},
	author = {Asch, Mark},
	doi = {10.1137/1.9781611976977},
	url = {https://epubs.siam.org/doi/abs/10.1137/1.9781611976977},
	publisher = {SIAM, Society for Industrial and Applied Mathematics},
	address = {Philadelphia, PA},
	year = {2022},
	doi = {10.1137/1.9781611976977},
}

@article{asch2018pathways,
    author = {M Asch and T Moore and R Badia and M Beck and P Beckman and T Bidot and F Bodin and F Cappello and A Choudhary and B de Supinski and E Deelman and J Dongarra and A Dubey and G Fox and H Fu and S Girona and W Gropp and M Heroux and Y Ishikawa and K Keahey and D Keyes and W Kramer and J-F Lavignon and Y Lu and S Matsuoka and B Mohr and D Reed and S Requena and J Saltz and T Schulthess and R Stevens and M Swany and A Szalay and W Tang and G Varoquaux and J-P Vilotte and R Wisniewski and Z Xu and I Zacharov},
    title ={Big data and extreme-scale computing: Pathways to Convergence-Toward a shaping strategy for a future software and data ecosystem for scientific inquiry},
    journal = {The International Journal of High Performance Computing Applications},
    volume = {32},
    number = {4},
    pages = {435-479},
    year = {2018},
    doi = {10.1177/1094342018778123},
    URL = { https://doi.org/10.1177/1094342018778123},
}

@book{luenberger1979,
    author = {David G. Luenberger},
    title = {Introduction to Dynamic Systems, Theory, Models, and Applications},
    publisher = {John Wiley \& Sons} ,
    year = {1979}
}

@book{friedland1985,
author = {Friedland, Bernard and Director, Stephen W.},
title = {Control Systems Design: An Introduction to State-Space Methods},
year = {1985},
isbn = {0070224412},
publisher = {McGraw-Hill Higher Education},
}

@book{powell2022, 
  title={Reinforcement learning and stochastic optimization: A unified framework for sequential decisions}, 
  author={Powell, Warren B.}, 
  year={2022}, 
  publisher={John Wiley \& Sons} 
}

@book{brucker2007, 
  title={Scheduling Algorithms}, 
  author={Peter Brucker}, 
  edition={{Fifth edition}}, 
  year={2007}, 
  publisher={Springer Berlin, Heidelberg} ,
  doi = "10.1007/978-3-540-69516-5"
}

@book{birgelouveaux2011, 
  title={Introduction to Stochastic Programming},
  doi={10.1007/978-1-4614-0237-4}, 
  author={John R. Birge and François Louveaux}, 
  edition={{Second edition}}, 
  year={2011}, 
  publisher={Springer New York, NY} ,
  doi = "10.1007/978-1-4614-0237-4"
}

@book{Pinedo2022, 
	title={Scheduling: Theory, Algorithms, and Systems}, 
	publisher={Springer Cham}, 
	author={Pinedo, Michael L.}, 
	edition={{6th edition}}, 
	year={2022}}

@book{shapiro_lectures_2009,
	title = {Lectures on {Stochastic} {Programming}},
	doi = {10.1137/1.9780898718751},
	url = {https://epubs.siam.org/doi/abs/10.1137/1.9780898718751},
	publisher = {SIAM, Society for Industrial and Applied Mathematics},
	author = {Shapiro, Alexander and Dentcheva, Darinka and Ruszczyński, Andrzej},
	year = {2009},
	doi = {10.1137/1.9780898718751}}

@article{Jian2015,
    author = {Jian Li and Aarif Mohamed Nazeer Batcha and Björn Gaining and Ulrich R. Mansmann},
    title ={An NGS Workflow Blueprint for DNA Sequencing Data and Its Application in Individualized Molecular Oncology},
    journal = {Cancer Informatics},
    volume = {14s5},
    number = {},
    pages = {CIN.S30793},
    year = {2015},
    doi = {10.4137/CIN.S30793},
    note ={PMID: 27081306},
    URL = {https://doi.org/10.4137/CIN.S30793},
}

@article{tanjo_practical_2021,
	title = {Practical guide for managing large-scale human genome data in research},
	volume = {66},
	issn = {1435-232X},
	url = {https://doi.org/10.1038/s10038-020-00862-1},
	doi = {10.1038/s10038-020-00862-1},
	number = {1},
	journal = {Journal of Human Genetics},
	author = {Tanjo, Tomoya and Kawai, Yosuke and Tokunaga, Katsushi and Ogasawara, Osamu and Nagasaki, Masao},
	month = jan,
	year = {2021},
	pages = {39--52},
}

@misc{MGG2025implementation,
    title = {Cognitive Continuum Digital Shadow---Implementation},
    author = {Marius Garénaux-Gruau and Olivier Martineau and François Bodin and Mark Asch},
    year = {2026},
    note = {In preparation},
    url={https://arxiv.org/abs/26xx.xxxx}, 
}

@misc{MGG2025usecases,
    title = {Cognitive Continuum Digital Shadow---Use Cases},
    author = {Marius Garénaux-Gruau and Mathis Certenais and François Bodin and Mark Asch},
    year = {2026},
    note = {In preparation},
    url={https://arxiv.org/abs/26xx.xxxx}, 
}

@book{bynum2021pyomo,
    title={Pyomo--optimization modeling in python}, 
    author={Bynum, Michael L. and Hackebeil, Gabriel A. and Hart, William E. and Laird, Carl D. and Nicholson, Bethany L. and Siirola, John D. and Watson, Jean-Paul and Woodruff, David L.}, 
    edition={Third}, 
    volume={67}, 
    year={2021}, 
    publisher={Springer Science \& Business Media} }

@article{hart2011pyomo, 
    title={Pyomo: modeling and solving mathematical programs in Python}, 
    author={Hart, William E and Watson, Jean-Paul and Woodruff, David L}, 
    journal={Mathematical Programming Computation}, 
    volume={3}, 
    number={3}, 
    pages={219--260}, 
    year={2011}, 
    publisher={Springer} }

@inproceedings{antypas_enabling_2021,
	address = {Orlando, FL, USA},
	title = {Enabling discovery data science through cross-facility workflows},
	copyright = {https://doi.org/10.15223/policy-029},
	isbn = {978-1-6654-3902-2},
	url = {https://ieeexplore.ieee.org/document/9671421/},
	doi = {10.1109/BigData52589.2021.9671421},
	language = {en},
	urldate = {2025-10-10},
	booktitle = {2021 {IEEE} {International} {Conference} on {Big} {Data} ({Big} {Data})},
	publisher = {IEEE},
	author = {Antypas, K. B. and Bard, D. J. and Blaschke, J. P. and Shane Canon, R. and Enders, Bjoern and Shankar, Mallikarjun Arjun and Somnath, Suhas and Stansberry, Dale and Uram, Thomas D. and Wilkinson, Sean R.},
	month = dec,
	year = {2021},
	pages = {3671--3680},
}

@book{Asch2026,
	title = {Optimal {Scheduling} for {Cross}-{Facility} {Workflows}},
	author = {Asch, Mark},
	url = {https://markasch.github.io/RCP4CDT/},
	year = {2026},
	publisher = {Online},
	note = {Executable companion guide; code at \url{https://github.com/markasch/RCPSP4CDT}},
}

@article{Unat2025,
	title = {The Persistent Challenge of Data Locality in the Post-Exascale Era},
	author = {Unat, Didem and Dubey, Anshu and Jeannot, Emmanuel and Shalf, John},
	journal = {Computing in Science \& Engineering},
	volume = {27},
	number = {4},
	pages = {19--27},
	year = {2025},
	doi = {10.1109/MCSE.2025.3567586},
}

@inproceedings{enders_superfacility_2020,
	title = {Cross-facility science with the {Superfacility} {Project} at {LBNL}},
	author = {Enders, Bjoern and Bard, Deborah and Snavely, Cory and Gerhardt, Lisa and Lee, Jason and Totzke, Becci and Antypas, Katie and Byna, Suren and Cheema, Ravi and Cholia, Shreyas and others},
	booktitle = {2nd {IEEE}/{ACM} Annual Workshop on Large-Scale Experiment-in-the-Loop Computing ({XLOOP})},
	pages = {1--7},
	year = {2020},
	doi = {10.1109/XLOOP51963.2020.00006},
}

@inproceedings{cruz_firecrest_2020,
	title = {{FirecREST}: a {RESTful} {API} to {HPC} systems},
	author = {Cruz, Felipe A. and Martinasso, Maxime and others},
	booktitle = {{IEEE}/{ACM} International Workshop on Interoperability of Supercomputing and Cloud Technologies ({SuperCompCloud})},
	pages = {21--26},
	year = {2020},
	doi = {10.1109/SuperCompCloud51944.2020.00009},
}

@book{Boyd2004,
	author = {Boyd, Stephen and Vandenberghe, Lieven},
	title = {Convex Optimization},
	publisher = {Cambridge University Press},
	year = {2004},
	doi = {10.1017/CBO9780511804441},
}

@book{Nocedal2006,
	author = {Nocedal, Jorge and Wright, Stephen J.},
	title = {Numerical Optimization},
	edition = {2},
	series = {Springer Series in Operations Research and Financial Engineering},
	publisher = {Springer},
	year = {2006},
	doi = {10.1007/978-0-387-40065-5},
}

\clearpage

\appendix

\section*{Appendix: Optimal schedule for the genomics workflow}

The model (1888 variables, 976 constraints, of which 208 binary) is solved to integer optimality by GLPK 5.0 in under 0.1\,s on a laptop. The expected total cost is \texteuro3\,995.53. The optimal first-stage decision is identical across scenarios (non-anticipativity): no job is launched at stage~1 and $J_1$ is hedged onto the reliable center $\mathrm{HP}_2$ at stage~2. Recourse then routes $J_2$, $J_3$, $J_4$ to the cheaper $\mathrm{HP}_1$, except in the persistent-outage scenario $S_4$ where $J_2$ remains on $\mathrm{HP}_2$.

\begin{table}[htbp]
\centering
\begin{tabular}{lccccc}
\hline
Scenario & $p_n$ & Execution & Transfer & Storage & Total (\texteuro) \\
\hline
$S_1$ (GG) & 0.36 & 700 & 32.0 & 3\,216.25 & 3\,948.25 \\
$S_2$ (GB) & 0.24 & 700 & 32.0 & 3\,216.25 & 3\,948.25 \\
$S_3$ (BG) & 0.24 & 700 & 32.0 & 3\,216.25 & 3\,948.25 \\
$S_4$ (BB) & 0.16 & 1\,000 & 27.5 & 3\,216.25 & 4\,243.75 \\
\hline
\multicolumn{5}{l}{Expected total cost} & 3\,995.53 \\
\hline
\end{tabular}
\caption{Per-scenario cost decomposition (\texteuro) of the optimal multistage stochastic schedule. Storage dominates ($\approx 81\%$ of total cost), an immediate, actionable insight for this workflow: reducing intermediate-data retention has far greater leverage than execution-site selection.}
\end{table}

\end{document}